\documentclass[amsmath,twocolumn,amssymb,prb,aps,superscriptaddress,letterpaper,showpacs]{revtex4}

\usepackage{graphicx}
\usepackage{dcolumn}
\usepackage{bm}
\usepackage{amsmath}

\begin{document}

\title{Spin decoherence of a heavy hole coupled to nuclear spins in a quantum dot}  %

\author{Jan Fischer}
\affiliation{Department of Physics, University of Basel, Klingelbergstrasse 82,
4056 Basel, Switzerland}

\author{W. A. Coish}
\affiliation{Department of Physics, University of Basel, Klingelbergstrasse 82,
4056 Basel, Switzerland}
\affiliation{Institute for Quantum Computing and Department of Physics and Astronomy,
University of Waterloo, 200 University Ave. W., Waterloo, ON, N2L 3G1,
Canada}

\author{D. V. Bulaev}
\affiliation{Department of Physics, University of Basel, Klingelbergstrasse 82,
4056 Basel, Switzerland}
\affiliation{Institute of Solid State Physics, Russian Academy of Sciences, 142432
Chernogolovka, Moscow District, Russia}

\author{Daniel Loss}
\affiliation{Department of Physics, University of Basel, Klingelbergstrasse 82,
4056 Basel, Switzerland}

\date{\today}

\begin{abstract}
We theoretically study the interaction of a heavy hole with nuclear spins in a
quasi-two-dimensional III-V semiconductor quantum dot and the resulting dephasing 
of heavy-hole spin states. It has frequently been stated in the literature that heavy 
holes have a negligible interaction with nuclear spins.
We show that this is not the case. In contrast, the interaction can be rather 
strong and will be the dominant source of decoherence in some cases.
We also show that for unstrained quantum dots the form of the interaction is 
Ising-like, resulting in unique and interesting 
decoherence properties, which might provide a crucial advantage to using dot-confined hole 
spins for quantum information processing, as compared to electron spins.
\end{abstract}


\maketitle

\section{Introduction}\label{sec:intro}

The spin of a quantum-dot-confined electron is considered a major candidate for
the realization of solid-state-based quantum bits, \cite{Loss1998} the basic building
blocks for quantum information processing devices and, eventually, for a quantum computer.
\cite{Cerletti2005}
One of the main obstacles to building these devices is the decay of spin coherence.
The ultimate limit to the coherence time for electron spins in most quantum 
dots at low temperatures is set by the hyperfine interaction with nuclei in the host material.
\cite{Burkard1999, Merkulov2002, Khaetskii2002, Coish2004}
If no special effort is made to control this environment, the associated coherence 
times are quite short, typically on the order of nanoseconds. 
\cite{Merkulov2002, Khaetskii2002, Coish2004, Petta2005}

Very recently, several experiments have shown initialization and readout of single hole 
spins in self-assembled quantum dots, \cite{Heiss2007, Ramsay2008, Gerardot2008,
Eble2008} 
and control over the number of holes in single gated quantum dots, \cite{Komijani2008}
prerequisites for single-hole-spin dephasing-time measurements.  
Ensemble hole-spin dephasing times have recently been measured in $p$-doped quantum wells. 
\cite{Syperek2007}
Hole-spin coherence times in III-V semiconductor quantum dots
are anticipated to be much longer than electron-spin coherence 
times due to a weak hyperfine coupling relative to conduction-band electrons.
\cite{Flissikowski2003, Shabaev2003, Woods2004, Bulaev2005a, Laurent2005,
Bulaev2007, Serebrennikov2007, Burkard2008}
In the present work, we show that, in contrast, the coupling 
of a heavy hole (HH) to the nuclear spins in a quantum dot can be rather strong, potentially
leading to coherence times that are comparable to those for electrons.
However, in the quasi-two-dimensional (Q2D) limit, this interaction takes-on a simple Ising-like
form, 
\begin{equation}
  \label{effham}
  H = \sum_k A^h_k \> s_z I_k^z, 
\end{equation}
where $A_k^h$ is the coupling of the HH to the $k^{\mathrm{th}}$ nucleus, 
$s_z$ is the hole pseudospin-$\frac{1}{2}$ operator, and $I_k^z$ is the $z$-component
of the $k^{\mathrm{th}}$ nuclear-spin operator $\mathbf{I}_k$. 
The form of this effective Hamiltonian has profound consequences for the
spin dynamics. Coherence times can be dramatically extended by 
preparing the slowly-varying nuclear field in a well-defined state (``narrowing''
the field distribution). \cite{Coish2004, Giedke2006, Stepanenko2006, 
Klauser2006, Reilly2008, Greilich2006, Greilich2007} 
For an electron spin interacting with nuclei via the 
contact hyperfine interaction, narrowing is effective only up to the 
time scale where slow internal nuclear-spin dynamics or 
transverse-coupling (``flip-flop'') terms become relevant.  Here we will 
show that heavy holes confined to two dimensions have negligible 
flip-flops, potentially leading to significantly longer spin coherence times.
The strong coupling of the HH to the nuclear spins is not due to
confinement but is also present in bulk crystals, while
the Ising-like interaction is a feature of Q2D systems.

This paper is organized as follows: In Sec. \ref{sec:interactions} we write down the
nuclear-spin interactions and derive an effective spin Hamiltonian for
a quantum-dot-confined HH.
In Sec. \ref{sec:decoherence} we calculate the dynamics of the transverse HH spin
for different external magnetic field directions. In Sec. \ref{sec:estimates}
we give estimates of the coupling strengths for the special case of a GaAs quantum
dot. Conclusions and comparison to recent experiments can be found in Sec. \ref{sec:conclusions}.
Technical details are deferred to Appendices \ref{appendix1}-\ref{appendix4}.

\section{Nuclear-spin interactions}\label{sec:interactions}

\subsection{Hamiltonians}

For a relativistic electron in the electromagnetic field of a nucleus with non-zero spin 
at position $\mathbf{R}_k$, there are three
terms that couple the electron spin and orbital angular momentum to the spin of the nucleus: 
the Fermi contact hyperfine interaction ($h^k_1$), a dipole-dipole-like interaction 
(the anisotropic hyperfine interaction, $h^k_2$), and the coupling of electron orbital angular 
momentum to the nuclear spin ($h^k_3$). Setting $\hbar = 1$, these interactions are described by
the following Hamiltonians: \cite{stoneham}
\begin{align}
  \label{ham:contact}
  h_1^k &= \frac{\mu_0}{4 \pi} \> \frac{8 \pi}{3} \> \gamma_S \gamma_{j_k} \>
  \delta(\mathbf{r}_{k}) \> \mathbf{S} \cdot \mathbf{I}_k,\\
  \label{ham:anisotropic}
  h_2^k &= \frac{\mu_0}{4 \pi} \> \gamma_S \gamma_{j_k} \> \frac{3 (\mathbf{n}_k \cdot 
  \mathbf{S}) (\mathbf{n}_k \cdot \mathbf{I}_k) - \mathbf{S} \cdot \mathbf{I}_k}
  {r_{k}^3 (1+d/r_{k})},\\
  \label{ham:angular}
  h_3^k &= \frac{\mu_0}{4 \pi} \> \gamma_S \gamma_{j_k} \> \frac{\mathbf{L}_k \cdot 
  \mathbf{I}_k} {r_{k}^3 (1+d/r_{k})}.
\end{align}
Here, $\gamma_S=2\mu_B$, $\gamma_{j_k}=g_{j_k} \mu_N$, $\mu_B$ is the Bohr magneton,
$g_{j_k}$ is the nuclear g-factor of isotopic species $j_k$,
$\mu_N$ is the nuclear magneton, $\mathbf{r}_{k} = \mathbf{r} - \mathbf{R}_k$ 
is the electron-spin position operator relative to the nucleus, 
$d \simeq Z \times 1.5 \times 10^{-15} \, \mathrm{m}$ is a length of nuclear dimensions, 
$Z$ is the charge of the nucleus, and $\mathbf{n}_k = \mathbf{r}_{k}/ r_{k}$. 
$\mathbf{S}$ and $\mathbf{L}_k = \mathbf{r}_k \times \mathbf{p}$ denote the spin and 
orbital angular-momentum operators of the electron, respectively.


Nuclear-spin interactions are typically much weaker than the spin-orbit interaction. 
It is therefore appropriate to form effective Hamiltonians with respect to
a basis of eigenstates of the Coulomb and spin-orbit interactions.
The $8 \times 8$ Kane Hamiltonian, which describes the band structure of a III-V 
semiconductor, provides such a basis. \cite{winkler, yucardona}
The Kane Hamiltonian is usually written in terms of 
conduction-band (CB) and valence-band (consisting of HH, light-hole (LH), 
and split-off sub-band) states. 
We derive an approximate basis of eigenstates in the HH sub-band
by projecting the $8 \times 8$ Kane Hamiltonian onto the two-dimensional HH subspace.

To form effective Hamiltonians, we must approximate the crystal-Hamiltonian eigenfunctions 
given by Bloch's theorem for a
single band $n$: $\Psi_{n \mathbf{k} \sigma}(\mathbf{r}) = \frac{1}{\sqrt{N_A}} e^{i \mathbf{k} 
\cdot \mathbf{r}} u_{n \mathbf{k} \sigma}(\mathbf{r})$, where $N_A$ is the number of atomic sites 
in the crystal, and the Bloch amplitudes $u_{n \mathbf{k} \sigma}(\mathbf{r})$ have the 
periodicity of the lattice.

We will approximate the $\mathbf{k}=\mathbf{0}$ Bloch amplitudes 
$u_{n \mathbf{0} \sigma}(\mathbf{r})$
within a primitive unit cell by a linear combination of atomic orbitals 
(see Eq. (\ref{blochHH}), below). Near an atomic site,
the CB Bloch amplitudes have approximate $s$-symmetry (angular momentum $l=0$), 
whereas the HH and LH Bloch amplitudes have approximate $p$-symmetry ($l=1$).
Adding spin, the $z$-component of total angular momentum of a HH is $m_J=\pm3/2$, 
whereas a LH has $m_J=\pm1/2$.
In the Q2D limit, i.e., going from the bulk crystal to a quantum well
(whose growth direction we take to be [001]), a splitting $\Delta_{\mathrm{LH}}$ develops between the 
HH and LH sub-bands at $\mathbf{k}=\mathbf{0}$.
We estimate $\Delta_{\mathrm{LH}} \simeq 100 \, \mathrm{m} e\mathrm{V}$ for a quantum well of height
$a_z \simeq 5 \, \mathrm{nm}$ 
in GaAs, much larger than the hyperfine coupling (see Appendix \ref{appendix1}).
The splitting $\Delta_{\mathrm{LH}}$ is essential since it produces a well-defined 
two-level system in the 
HH sub-band, and we can restrict our considerations to the manifold
of $m_J=\pm3/2$ states.

\subsection{Interactions in an atom}

Before addressing confinement in quantum dots, we illustrate that the interaction of 
an electron in a hydrogenic $p$ orbital with the spin of the nucleus 
(which we choose to be at $\mathbf{R}_k=\mathbf{0}$) is generally non-zero. Moreover, when
projected onto the manifold of $m_J=\pm3/2$ states, this interaction takes-on a simple
Ising form. 
Although our final analysis will apply to any III-V semiconductor, we will take GaAs as a concrete
example.

The effective screened  nuclear charges $Z_\mathrm{eff}$ ``felt'' by the valence electrons 
(in $4s$ and $4p$ orbitals) in Ga and As atoms have been calculated in Ref. 
\onlinecite{Clementi1963}.
The $4s$ orbitals and $4p$ orbitals (with orbital angular momentum $m_L=\pm1$) are given in 
terms of hydrogenic eigenfunctions with the replacements $Z\to Z_\mathrm{eff}$ by
$\Psi_{400}(\mathbf{r}) = R_{40}(r) Y_0^0(\theta, \varphi)$ and
$\Psi_{41\pm1}(\mathbf{r}) = R_{41}(r) Y_1^{\pm1}(\theta, \varphi)$, respectively.
Including spin and evaluating matrix elements of the Hamiltonians 
(\ref{ham:contact})-(\ref{ham:angular})
with respect to hydrogenic $4s$ states leads to effective spin Hamiltonians of the form
\cite{stoneham, abragam}
$h_1^{4s} = A_s \mathbf{S} \cdot \mathbf{I}_k$ and, due to the spherical symmetry of the 
wavefunction,  $h_2^{4s} = h_3^{4s} = 0$.
The same procedure with the $4p$ states leads to effective Hamiltonians $h_1^{4p} = 0$ 
(since $p$-states vanish at
the origin) and $h_2^{4p} + h_3^{4p} = A_p s_z I_k^z$, where $s_z=\pm\frac{1}{2}$ 
corresponds to $m_J=\pm3/2$.
$A_s$ and $A_p$ denote the coupling strengths of electrons in $4s$ and $4p$ 
orbitals, respectively. Evaluating all integrals exactly gives
\begin{equation}
  \label{Aratio}
  \frac{A_p}{A_s} = \frac{1}{5} 
\left( \frac{Z_\mathrm{eff}(\kappa,4p)}{Z_\mathrm{eff}(\kappa,4s)} 
  \right)^3,\;\kappa=\mathrm{Ga},\,\mathrm{As}.
\end{equation}

Quite significantly, after inserting values of $Z_\mathrm{eff}$ from 
Ref. \onlinecite{Clementi1963} into Eq. (\ref{Aratio}), we find that the ratio of coupling
strengths is fairly large -- on the order of 10\%:
$A_p/A_s \simeq 0.14$ for Ga, and $A_p/A_s \simeq 0.11$ for As.
Since $h_2^k$ and $h_3^k$ do not contribute to the hyperfine interaction of an electron
in an $s$ orbital, research on hyperfine interaction for electrons in an $s$-type
conduction band has focused on the contact term 
$h_1^k$, neglecting the other interactions. The fact that $h_1^k$, in contrast, gives no
contribution for an electron in a $p$ orbital has led to the claim that 
electrons in $p$ orbitals (and holes) do not interact with nuclear spins. 
Eq. (\ref{Aratio}) shows that 
$h_2^k$ and $h_3^k$ can contribute significantly.
Furthermore, while the interaction of a $4s$-electron with the nuclear spin is of Heisenberg type, 
within the manifold of $m_J=\pm3/2$ states,
the interaction of a $4p$-electron is Ising-like at leading order (virtual transitions via the 
$m_J=\pm1/2$ states may lead to non-Ising corrections). This result is a direct consequence 
of the Wigner-Eckart theorem and the fact that the Hamiltonians 
(\ref{ham:contact})-(\ref{ham:angular}) can be written in the form 
$h_i^k = \mathbf{v}_i^k \cdot \mathbf{I}_k$, 
where $\mathbf{v}_i^k$ are vector operators in the electron (spin and orbital) Hilbert space.

\subsection{Interactions in a quantum dot}

We now return to the problem directly relevant to a HH in a two-dimensional quantum well.
We consider an additional circular-symmetric parabolic confining potential in the plane
of the quantum well defining a quantum dot. 
Neglecting hybridization with other bands, which we estimate to be typically on the order 
of 1\% (see Appendix \ref{appendix1}), the pseudospin states for a HH within the envelope-function 
approximation read
$|\Psi_{\sigma}\rangle = |\phi; u_{\mathrm{HH} \mathbf{0} \sigma} \rangle |\sigma \rangle$,
where $|{\phi; u_{\mathrm{HH} \mathbf{0} \sigma}} \rangle$ and $|{\sigma} \rangle$ ($\sigma=\pm$)
denote the orbital and spin states, respectively. 
The orbital wavefunctions are given explicitly by
$\langle \mathbf{r} | \phi; u_{\mathrm{HH} \mathbf{0} \sigma} \rangle = \sqrt{v_0} \> \phi(\mathbf{r})
u_{\mathrm{HH} \mathbf{0} \sigma}(\mathbf{r})$. Here, $v_0$ is the volume occupied by a single atom
(half the volume of a two-atom zinc-blende primitive unit cell), 
and $\phi(\mathbf{r})=\phi_z(z)\phi_\rho(\rho)$ is the envelope function.
The radial ground-state envelope function is a Gaussian:
\begin{equation}
  \label{envelope}
  \phi_\rho(\rho) = \frac{1}{\sqrt{\pi} l} \> \exp \left( -\frac{\rho^2}{2 l^2} \right),
\end{equation}
where $\pmb{\rho}=(x,y)$, $\rho=|\pmb{\rho}|$, $l=l_0 \, [1+(B_z/B_0)^2]^{-1/4}$, $B_z$ is 
the component of an externally applied magnetic field along 
the growth direction and $B_0=\Phi_0/\pi l_0^2$ where $\Phi_0=h/|e|$ is a flux quantum.  
A typical dot Bohr radius of $l_0=30\,\mathrm{nm}$ gives $B_0 \simeq 1.5\,\mathrm{T}$.

In a solid, the HH is delocalized over the lattice sites of the crystal.
The nuclei do not interact solely with the fraction of the HH in the same primitive 
unit cell (`on-site' interaction), but also with density localized at more distant
atomic sites (long-ranged interactions). We neglect the long-ranged interactions,  
which lead to corrections on the order of 1\% relative to the on-site interaction 
(see Appendix \ref{appendix3}). 
If the envelope function varies slowly on the length scale of a primitive cell,  
we find (combining $h^k_2$ and $h^k_3$) $A_k^h = A_h^{j_k} v_0 |\phi(\mathbf{R}_k)|^2$, where
\begin{equation}
  \label{holecoupling}
  A_h^{j_k} = -\frac{\mu_0}{4 \pi} \> \gamma_S \gamma_{j_k} \left\langle \frac{3 \cos^2 \theta_{k} + 1}
  {r_{k}^3(1+d/r_{k})} \right\rangle_{\mathrm{p.c.}}.
\end{equation}
Here, $\langle \cdots \rangle_{\mathrm{p.c.}}$ denotes the expectation value with respect to 
$u_{\mathrm{HH} \mathbf{0} \sigma}(\mathbf{r})$ over a primitive unit cell
($\sigma=\pm$ give the same matrix elements), and $\theta_k$ is the polar angle of 
$\mathbf{r}_k$.
The magnetic moment of a HH is inverted with respect to that for an electron. 
This results in a change of sign in Eqs. 
(\ref{ham:contact})-(\ref{ham:angular}) and leads to the minus sign in 
Eq. (\ref{holecoupling}) and of the values in columns (i) and (ii) of Table \ref{table1}.

\subsection{Non-Ising corrections}
There will be small corrections to the form of the effective Hamiltonian given in Eq.
(\ref{effham}).
Evaluating off-diagonal matrix elements of \eqref{ham:anisotropic} and \eqref{ham:angular} 
with the approximate Bloch amplitudes \eqref{blochHH} yields non-Ising terms, whose
associated coupling strengths $A_h^\perp$ we find to be small: $A_h^\perp < 0.06 \, A_h^{j}$.
Higher-order virtual transitions between the $m_J=\pm3/2$ states via the LH sub-band
are suppressed by $\sim A_k^h/\Delta_{\mathrm{LH}} \ll 1$.
Hybridization with other bands can also lead to non-Ising corrections. 
For unstrained quantum dots, we find that these corrections are small:
typically on the order of 1\% of the values given in Table I (see also
Appendix \ref{appendix1}). Strain can lead to considerably stronger band mixing and,
hence, to significantly larger non-Ising corrections to Eq. \eqref{effham}.

\section{Spin Decoherence}\label{sec:decoherence}

\begin{figure}
\centering
\includegraphics[width=\columnwidth]{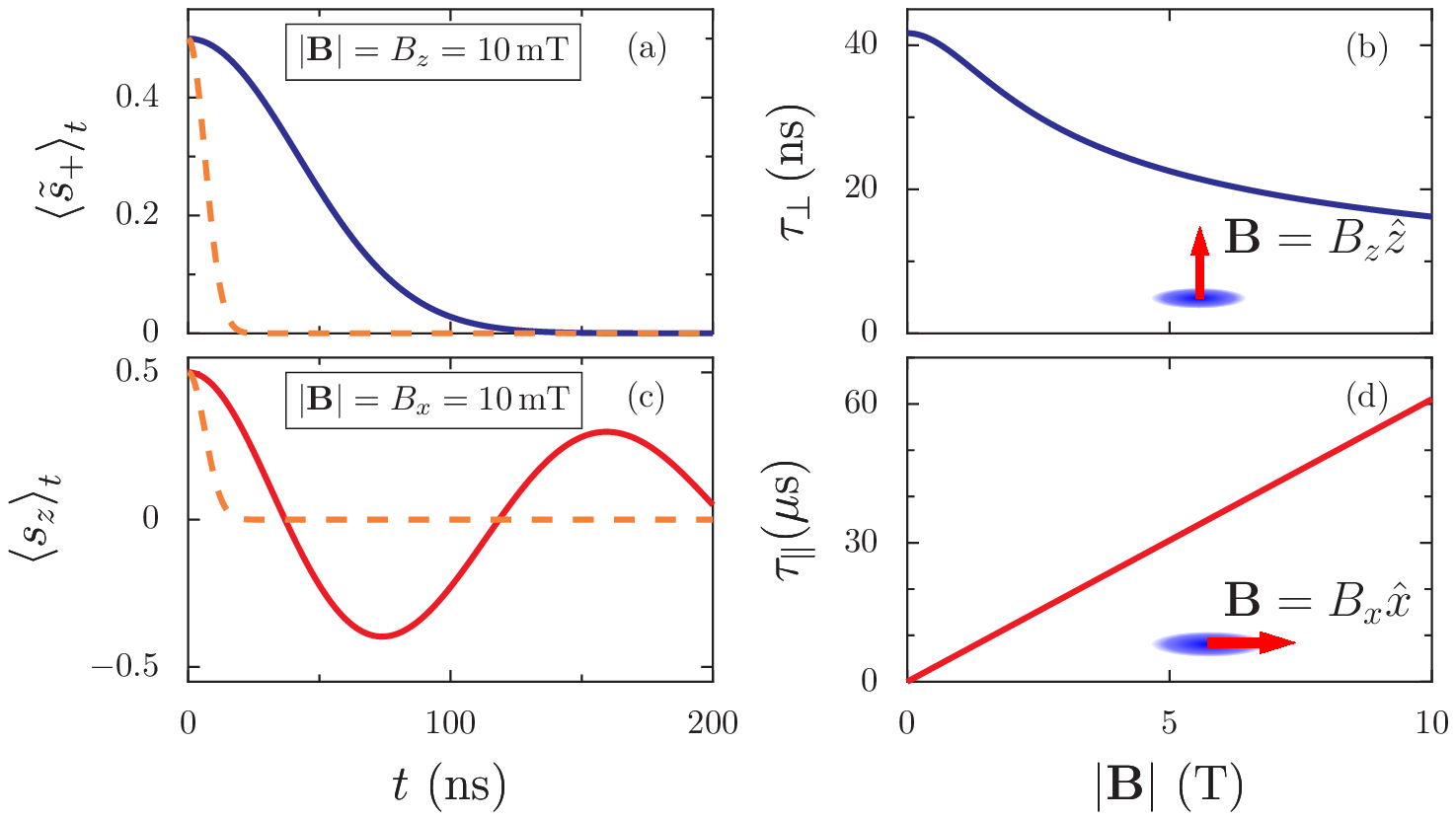}
\caption{\label{fig:decay}Dephasing of HH pseudospin states (solid lines in (a) and (c)).  
  The decay is Gaussian for an out-of-plane magnetic field $B_z$
  (a) (see Eq. (\ref{eq:splus})) and given by a slow power law at long times ($\sim 1/\sqrt{t}$) 
  for an in-plane magnetic field $B_x$ (c) (see Eq.(\ref{eq:sz})).  
  We have chosen $|\mathbf{B}|=10 \, \mathrm{mT}$ in (a) and (c).
  Magnetic-field dependences of the relevant coherence times are shown for 
  a field that is out-of-plane (b) and in-plane (d).  
  We have assumed $g_\parallel=0.04$ (from Ref. \onlinecite{Marie1999})
  and a zero-field lateral dot size $l_0=30\,\mathrm{nm}$ and 
  height $a_z=5\,\mathrm{nm}$, leading to $N=\pi l_0^2a_z/v_0=6.5\times 10^5$ nuclei within the dot 
  at $B_z=0$. We have taken $v_0 = a_L^3/8$, where $a_L = 5.65 \, \mathrm{\AA}$ 
  is the GaAs lattice constant. The dashed lines in (a) and (c)
  show the dephasing of CB electron spin states in the high-field limit,
  $|\mathbf{B}| \gg \sigma_e/g_e \mu_B$, where $g_e$ is the electron g-factor, and $\sigma_e$ is
  obtained from Eq. (\ref{sigma}) by replacing $A_h^j$ by the electron hyperfine coupling 
  constants $A_e^j$.}
\end{figure}

Now that the effective Hamiltonian (Eq. (\ref{effham})) has been established, we can analyze 
the dephasing of a HH pseudospin in the presence of a random nuclear environment.  
In an applied magnetic field, pseudospin dynamics of the HH are described by the Hamiltonian  
\begin{equation}
\label{hamwithfields}
H=\left(b_\perp +h_z\right)s_z+b_\parallel s_x,
\end{equation}
where $h_z=\sum_k A_k^h I_k^z$ is the nuclear field operator, $b_\perp=g_\perp\mu_B B_z$ is the Zeeman splitting due to a magnetic field $B_z$ along the growth direction, and $b_\parallel=g_\parallel \mu_B B_x$ is the Zeeman splitting due to an applied magnetic field $B_x$ in the plane of the quantum dot.  $g_\perp$ and $g_\parallel$ are the components of the HH g-tensor along the growth direction and in the plane of the quantum dot, respectively (we assume the in-plane g-tensor to be isotropic).

If no special effort is made to control the nuclear field, the field value will be Gaussian-distributed in the limit of a large number of nuclear spins. \cite{Coish2004}  The variance for a random 
nuclear-spin distribution is (see Appendix \ref{appendix4})
\begin{equation}
  \label{sigma}
  \left<h_z^2\right>=\sigma^2 \simeq \frac{1}{4 N}
  \sum_j\nu_jI^j(I^j+1)(A^j_h)^2,
\end{equation} 
where $N=\pi l^2a_z/v_0$ is the number of nuclei within the quantum dot.  The nuclear-field fluctuation $\sigma$ therefore inherits a magnetic-field dependence from $l$ (see Eq. (\ref{envelope})).  A finite nuclear-field variance will result in a random distribution of precession frequencies experienced by the hole pseudospin, inducing pure dephasing (decay of the components of hole pseudospin transverse to 
$\mathbf{B}$).

First, we consider the case $b_\parallel=0$. For a hole pseudospin initially oriented along the $x$ direction, we find a Gaussian decay (see {Fig. \ref{fig:decay} (a)})
of the transverse pseudospin in the rotating frame 
$\left<\tilde{s}_+\right>_t=\exp(-ib_\perp t)\left(\left<s_x\right>_t+i\left<s_y\right>_t\right)$:
\begin{equation}
  \label{eq:splus}
  \left<\tilde{s}_+\right>_t=\frac{1}{2}\exp{\left(-\frac{t^2}{2 \tau_\perp^2}\right)},
  \quad \tau_\perp = \frac{1}{\sigma}.
\end{equation} 
This is the same Gaussian decay that occurs for electrons. \cite{Merkulov2002, Khaetskii2002,
Coish2004} Here, since the magnetic field is taken to be out-of-plane, we must take account 
of the diamagnetic ``squeezing'' of the wavefunction. This squeezing affects the number $N$ of nuclear spins within the dot and hence, the finite-size fluctuation $\sigma$.  The coherence time $\tau_\perp(B_z)=1/\sigma(B_z)=\tau_\perp(0)\left[1+(B_z/B_0)^2\right]^{-1/4}$ then \emph{decreases} for large $B_z$ (see {Fig. \ref{fig:decay} (b)}). This undesirable effect can be avoided for confined electron spins by generating a large Zeeman splitting through an in-plane (rather than out-of-plane) magnetic field.  This option may not be available for a HH where, typically, $g_\parallel\ll g_\perp$.

The situation changes drastically for an in-plane magnetic field ($b_\perp=0$).  
In this case, since the hyperfine fluctuations are purely transverse to the applied field direction, 
the decay is given by a slow power law at long times (see {Fig. \ref{fig:decay} (c)})
and the relevant dephasing time 
\emph{increases} as a function of the applied magnetic field (see {Fig. \ref{fig:decay} (d)}).
In the limit $b_\parallel\gg \sigma$ and for a HH pseudospin initially prepared along the $\hat{z}$-direction, we find
\begin{equation}
  \label{eq:sz}
  \left<s_z\right>_t\simeq\frac{\cos\left(b_\parallel t+\frac{1}{2}\arctan(t/\tau_\parallel)\right)}
  {2[1+(\frac{t}{\tau_\parallel})^2]^{1/4}}, \quad \tau_\parallel=\frac{b_\parallel}{\sigma^2}.
\end{equation}
The derivation of Eq. (\ref{eq:sz}) is directly analogous to that for the decay of driven 
Rabi oscillations in Ref. \onlinecite{Koppens2007}.

\section{Estimates of the coupling strengths}\label{sec:estimates}

\begin{table}[t]
  \centering
  \begin{tabular}{|c|c|c||c|c|}
    \hline
    j & \multicolumn{2}{c||}{$A_h^j$ ($\mu e \mathrm{V}$)} & 
    \multicolumn{2}{c|}{$A_e^j$ ($\mu e \mathrm{V}$)}\\
    \hline
    & (i) & (ii) & (iii) & (iv)\\
    \hline
    ${}^{69}\mathrm{Ga}$ & -7.1 & -13 & 40 & 74\\
    ${}^{71}\mathrm{Ga}$ & -9.0 & -17 & 51 & 94\\
    ${}^{75}\mathrm{As}$ & -8.2 & -12 & 59 & 89\\
    \hline
    $A_\alpha = \sum_j \nu_j A_\alpha^j \quad (\alpha=e,h)$ & -8.0 & -13 & 52 & 86\\
    \hline
\end{tabular}
  \caption{Estimates of the coupling strengths for a HH ($A_h^j$) and a CB electron ($A_e^j$) 
    for the three isotopes in GaAs. 
    Columns (i) and (iii) show values obtained from a linear combination of hydrogenic 
    eigenfunctions, using free-atom values of $Z_\mathrm{eff}$ calculated in
    Ref. \onlinecite{Clementi1963}.
    Column (iv) gives the accepted values of $A_e^j$ from Ref. \onlinecite{Paget1977}. Column (ii)
    shows the rescaled values from column (i) (see text). In the last row we give the 
    average coupling constants weighted by the natural isotopic abundances: 
    $\nu_{^{69}\mathrm{Ga}}=0.3$, $\nu_{^{71}\mathrm{Ga}}=0.2$, $\nu_{^{75}\mathrm{As}}=0.5$.}
    \label{table1}
\end{table}

To estimate the size of $A_h^j$, we need an explicit expression for
the HH ${\mathbf{k}=\mathbf{0}}$ Bloch amplitudes. We approximate  
$u_{\mathrm{HH} \mathbf{0} \sigma}(\mathbf{r})$ within a Wigner-Seitz cell centered halfway along 
the Ga-As bond by a linear combination of atomic orbitals, following
Ref. \onlinecite{Gueron1964}:
\begin{align}
  u_{\mathrm{HH} \mathbf{0} \sigma}(\mathbf{r}) \Bigr|_{\mathbf{r} \in \mathrm{WS}}
  = &N_{\alpha_v} \Bigl( \alpha_v \Psi_{41\sigma}^{\mathrm{Ga}}
  (\mathbf{r}+\mathbf{d}/2) \nonumber \\
  &+ \sqrt{1-\alpha_v^2} \Psi_{41\sigma}^{\mathrm{As}} (\mathbf{r}-\mathbf{d}/2) \Bigr).
  \label{blochHH}
\end{align}
Here, $\mathbf{d}=\frac{a}{4} (1,1,1)$ is the Ga-As bond vector, $a$ is the lattice constant, 
$\alpha_v$ describes the relative electron sharing at the Ga and As sites
in the HH sub-band, and $N_{\alpha_v}$ is a normalization constant, chosen to enforce
$\int_{\mathrm{WS}} d^3 r \> |u_{\mathrm{HH} \mathbf{0} \sigma}(\mathbf{r})|^2 = 2$,
where the integration is performed over the Wigner-Seitz cell defined above.
To simplify numerical integration, we replace the 
Wigner-Seitz cell by a sphere centered halfway along the Ga-As bond, with radius 
given by half the Ga-Ga nearest-neighbor distance.
We find the electron sharing in the CB from the densities given in Ref. \onlinecite{Paget1977} 
to be $\alpha_c^2 \simeq 1/2$ (see Appendix \ref{appendix2}) and assume the same
($\alpha_v^2=1/2$) for Eq. (\ref{blochHH}).
Using $Z_\mathrm{eff}$ for free atoms, \cite{Clementi1963}
we evaluate Eq. (\ref{holecoupling}) with the ansatz, Eq. (\ref{blochHH}), 
by numerical integration, giving the values shown in Table \ref{table1}, column (i).
We check the validity of this procedure by writing the CB Bloch amplitudes as in Eq. 
(\ref{blochHH}), replacing the $4p$-eigenfunctions by $4s$-eigenfunctions.
Evaluating the coupling constants for the CB ($A_e^j$) from $h_1^k$
gives the numbers in column (iii). The accepted values of $A_e^j$ from
Ref. \onlinecite{Paget1977} are shown in
column (iv) for comparison. Our method produces $A_e^j$ to within a factor of two
of the accepted values.
Our procedure, which relies on free-atom orbitals, most likely under-estimates the 
electron density near the atomic sites, which should be enhanced in a solid due to 
confinement.  Assuming the relative change in density going from a free atom to a solid 
is the same for the CB and HH band, we rescale
the results in column (i) by the ratio of the values in columns (iv) and (iii),
giving column (ii). 
Due to the approximations involved, we expect the values in columns (i) and (ii) only
to be valid to within a factor of two or three.

\section{Conclusions}\label{sec:conclusions}

We have shown that the interaction of a quantum-dot-confined heavy hole with
nuclear spins is stronger than previously anticipated. We have estimated the associated 
coupling strength to be on the order of $10 \mu e \mathrm{V}$ in GaAs -- only one order 
of magnitude less than the hyperfine coupling for electrons. However, the 
interaction turns out to be Ising-like which
has profound consequences for hole-spin decoherence. Since no flip-flop terms occur
in the effective Hamiltonian (Eq. \eqref{effham}), the main source of decoherence is given
by the broad frequency distribution of the nuclear spins. Recent theoretical
and experimental studies have shown that state-narrowing techniques are capable of
strongly suppressing this source of decoherence, which makes the heavy hole an
attractive spin-qubit candidate.

Very recently, experimental results on hole-spin relaxation in self-assembled quantum 
dots have been released. \cite{Gerardot2008, Eble2008}
Gerardot \textit{et al.} report an extremely weak coupling of HH spin states,
which is explained by our theory to be a direct consequence of the Ising-like nuclear-spin
interaction (negligible flip-flop terms). 
Eble \textit{et al.}, in contrast, find very short hole-spin relaxation
times on the order of 15 nanoseconds. This is due to the
strong strain present in the particular dots used in this experiment, resulting 
in a considerable HH-LH mixing and a highly non-Ising interaction (large
flip-flop terms).

\begin{acknowledgements}
We acknowledge discussions with S. I. Erlingsson and D. Stepanenko, and 
funding from the Swiss NSF, NCCR Nanoscience, JST ICORP, QuantumWorks,
an Ontario PDF (WAC), and the ``Dynasty'' Foundation (DVB).
\end{acknowledgements}

\begin{appendix}

\section{Heavy-hole states}
\label{appendix1}

In this section we give details on our derivation of an approximate basis of heavy-hole (HH) 
eigenstates in a quantum dot. We will approximate the ground-state
quantum-dot envelope function in the HH sub-band by
\begin{align}
  \label{supp:envelope}
  \phi(\mathbf{r}) &= \phi_{z}(z) \, \phi_{\rho}(\rho),\\
  \label{supp:enveloperho}
  \phi_{\rho}(\mathbf{\rho}) &= \frac{1}{\sqrt{\pi}l}\exp\left(-\frac{\rho^{2}}{2l^{2}}\right),\\
  \label{supp:envelopez}
  \phi_{z}(z) &= \sqrt{\frac{2}{a_{z}}}\sin\left(\frac{\pi z}{a_{z}}\right),
  \quad z=[0\ldots a_{z}],
\end{align}
where $a_z$ is the width of the confinement potential along the growth direction
(for definition of the other symbols see Eq. (\ref{envelope})).
We will then estimate the size of the splitting 
$\Delta_{\mathrm{LH}}$ between the HH and the light-hole (LH) band and the degree of 
hybridization with the conduction band (CB), LH and split-off (SO) sub-bands.

We start from the $8 \times 8$ Kane Hamiltonian given in Ref. \onlinecite{winkler:appendix} for
bulk zinc-blende-type crystals, which is written in terms of the exact eigenstates 
(near $\mathbf{k}=\mathbf{0}$) of an 
electron in the CB, HH, LH and SO band, usually denoted by 
$|1/2; \pm 1/2 \rangle_c$, $|3/2; \pm 3/2 \rangle_v$, $|3/2; \pm 1/2 \rangle_v$,
and $|1/2; \pm 1/2 \rangle_v$, respectively. We neglect terms that are more than two orders
of magnitude smaller than the fundamental band-gap energy $E_g$
\footnote{In Winkler's notation, \cite{winkler:appendix} these are the terms proportional to
$C$, $B_{7v}$, and $B_{8v}^\pm$. These terms will not lead to considerable
corrections for our purposes. However, the terms $C k_\pm$ could become
relevant when considering higher-order effects in the spin-orbit interaction, 
such as the cubic Dresselhaus terms which were derived in Ref. \onlinecite{Bulaev2005a}.}
and perform the quasi-two-dimensional limit
by assuming that a confinement potential has been applied along the growth 
direction. If the confinement potential is sufficiently strong (i.e., if the quantum well
is sufficiently narrow), the energy-level spacing
will be large and the electron will be in the ground state at low temperatures.
Any operator acting on the $z$-component of the electron envelope function can then be replaced
by its expectation value with respect to the $z$-component of the ground-state envelope function.
For the Kane Hamiltonian this means that we can replace powers of the $z$-component $\hbar k_z$ 
of the crystal momentum $\hbar  \mathbf{k}$ by expectation values. 
Assuming an infinite square-well potential
of width $a_z$ confining the electron along the growth direction, the ground state is given by
Eq. (\ref{supp:envelopez}).
Calculating the expectation value of $k_z$ and $k_z^2$ with respect to the ground state,
we find $\langle k_z \rangle = 0$ and $\langle k_z^2 \rangle = \pi^2/a_z^2$. 
This allows us to write the Kane Hamiltonian in the following form:
\begin{equation}
  \label{supp:kane}
  H_K = 
  \left( \begin{array}{cccc}
      H_{\mathrm{CB}} & V_1 & V_2 & V_3\\
      V_1^{\dagger} & H_{\mathrm{HH}} & V_4 & V_5\\
      V_2^{\dagger} & V_4^{\dagger} & H_{\mathrm{LH}} & V_6\\
      V_3^{\dagger} & V_5^{\dagger} & V_6^{\dagger} & H_{\mathrm{SO}}
    \end{array} \right),
\end{equation}
where 
\begin{equation*}
  \begin{array}{ll}
  H_{\mathrm{CB}} = \begin{pmatrix} A & 0\\ 0 & A\end{pmatrix},&
  H_{\mathrm{HH}} = \begin{pmatrix} B & 0\\ 0 & B\end{pmatrix},\\
  H_{\mathrm{LH}} = \begin{pmatrix} C & 0\\ 0 & C\end{pmatrix},&
  H_{\mathrm{SO}} = \begin{pmatrix} D & 0\\ 0 & D\end{pmatrix},\\
  V_1 = \frac{1}{\sqrt{2}} \begin{pmatrix} -E & 0\\ 0 & E^*\end{pmatrix},&
  V_2 = \frac{1}{\sqrt{6}} \begin{pmatrix} 0 & E^*\\ -E & 0\end{pmatrix},\\
  V_3 = \frac{1}{\sqrt{3}} \begin{pmatrix} 0 & -E^*\\ -E & 0\end{pmatrix},&
  V_4 = \sqrt{3} \begin{pmatrix} 0 & F\\ F^* & 0\end{pmatrix},\\
  V_5 = \sqrt{6} \begin{pmatrix} 0 & -F\\ F^* & 0\end{pmatrix},&
  V_6 = \sqrt{2} \begin{pmatrix} -G & 0\\ 0 & G \end{pmatrix},
  \end{array}
\end{equation*}
and
\begin{align*}
  A &= E_g + \hbar^2 (k_x^2 + k_y^2 + \langle k_z^2 \rangle)/2m',\\
  B &= -\epsilon [(\gamma'_1 + \gamma'_2) (k_x^2 + k_y^2) + (\gamma'_1 - 2\gamma'_2) 
    \langle k_z^2 \rangle],\\
  C &= -\epsilon [(\gamma'_1 - \gamma'_2) (k_x^2 + k_y^2) + (\gamma'_1 + 2\gamma'_2) 
    \langle k_z^2 \rangle],\\
  D &= -\epsilon \gamma'_1 (k_x^2 + k_y^2 + \langle k_z^2 \rangle) - \Delta_{\mathrm{SO}},\\
  E &= P k_+,\\
  F &= \epsilon [\gamma'_2 (k_x^2 - k_y^2) -2i \gamma'_3 k_x k_y ],\\
  G &= \epsilon \gamma'_2 (k_x^2 + k_y^2 - 2 \langle k_z^2 \rangle).
\end{align*}
Here, $\epsilon = \hbar^2/2m_0$ and $m_0$ is the free-electron mass,
whereas $m'$ is the effective mass of a CB electron. Furthermore, ${k_\pm = k_x \pm i k_y}$,
$\gamma'_j$ denote the Luttinger parameters, $P$ is the inter-band momentum,
and $\Delta_{\mathrm{SO}}$ is the spin-orbit 
gap between the LH and the SO bands. Experimental values for these
parameters can be found in Table \ref{supp:table}.

We assume a circular-symmetric parabolic confinement potential with frequency $\omega_0$
in the $xy$-plane defining a quantum dot.
Including a magnetic field along the growth direction, the ground state is approximately 
described by the Gaussian given in Eq. (\ref{supp:enveloperho}).
The envelope function of the quantum dot is then the product of the in-plane and out-of-plane 
components, as given in Eq. (\ref{supp:envelope}).

In the quasi-two-dimensional limit, a gap 
\begin{equation}
  \label{supp:deltaLH}
  \Delta_{\mathrm{LH}} = \langle B-C \rangle = -\frac{\hbar^2 \gamma'_2}{m_0} \Bigl( \langle k_x^2 \rangle
  + \langle k_y^2 \rangle - 2 \langle k_z^2 \rangle \Bigr)
\end{equation}
develops between the HH and LH sub-bands, lifting the HH-LH degeneracy.
Here, $\langle \cdots \rangle$ denotes the expectation value with respect to (\ref{supp:envelope}).
The in-plane level spacing scales like $\sim 1/l^2$, where $l$ is the dot Bohr radius.
The in-plane level spacing is much smaller than the level spacing along 
the growth direction since, for typical dots,
$a_z^2 \ll l^2$. Neglecting $\langle k_x^2 \rangle$ and $\langle k_y^2 \rangle$ compared to 
$\langle k_z^2 \rangle$ in Eq. (\ref{supp:deltaLH}) and inserting $\langle k_z^2 \rangle 
= \pi^2/a_z^2$ for a square-well potential, we estimate
\begin{equation}
  \Delta_{\mathrm{LH}} \simeq \frac{2 \pi^2 \gamma'_2 \hbar^2}{a_z^2 m_0}\
  \simeq 100 \, \mathrm{m}e\mathrm{V}
\end{equation}
for $a_z \simeq 5 \, \mathrm{nm}$, using $\gamma'_2 \simeq 2.06$ 
for GaAs (see Table \ref{supp:table}).
The HH-LH splitting is thus much larger than the typical energy scale associated with 
the hyperfine interaction ($A_e \simeq 90 \, \mu e \mathrm{V}$ for CB electrons in GaAs).

To derive the approximate electron eigenfunctions in the HH sub-band of the quantum well,
we start from the Kane Hamiltonian (\ref{supp:kane}).
We use quasi-degenerate perturbation theory up to first order in $1/\mathcal{E}$
(where $\mathcal{E}$ stands for $E_g$, $\Delta_{\mathrm{LH}}$, or
$\Delta_{\mathrm{LH}}+\Delta_{\mathrm{SO}}$), taking $H_{\mathrm{HH}}$ as the unperturbed Hamiltonian.
\cite{winkler:appendix} This leads to a band-hybridized state of the form 
\begin{align}
  \label{supp:hybridization}
  |\Psi^\sigma_{\mathrm{HH,hyb}} \rangle = \mathcal{N}_\sigma \sum_n \lambda_n^\sigma 
  |\phi_{n \sigma}; u_{n \mathbf{0} \sigma}\rangle.
\end{align}
Here, $\sigma = \pm$, $\langle \mathbf{r} | \phi_{n \sigma}; u_{n \mathbf{0} \sigma} \rangle = 
\sqrt{v_0} \phi_{n \sigma}(\mathbf{r}) \, u_{n \mathbf{0} \sigma} (\mathbf{r})$ is the product of envelope
function and $\mathbf{k}=\mathbf{0}$ Bloch amplitude in band $n$ 
(CB, HH, LH or SO), the prefactors $\lambda_n^\sigma$ describe the degree of band hybridization, 
and $\mathcal{N}_\sigma$ enforces proper normalization.

In first order quasi-degenerate perturbation theory, the hybridization with the CB and
the LH and SO sub-bands is described by the interaction terms $V_1$, $V_4$, and $V_5$
in Eq. (\ref{supp:kane}), respectively.
We estimate the degree of hybridization by applying these operators to a two-spinor
containing the in-plane ground-state envelope function (\ref{supp:enveloperho}) of the 
HH sub-band. For the hybridization with the conduction band, we find 
(for $\mathbf{B}=\mathbf{0}$)
\begin{equation}
  -\frac{1}{E_g} V_1 \begin{pmatrix} \phi_\rho(\rho)\\ \phi_\rho(\rho) \end{pmatrix} =
  \begin{pmatrix} \lambda_{\mathrm{CB}}^+ \, \phi_{\mathrm{CB} \rho +}(\rho)\\ \lambda_{\mathrm{CB}}^-
  \, \phi_{\mathrm{CB} \rho -}(\rho) \end{pmatrix},
\end{equation}
where
\begin{equation}
  \phi_{\mathrm{CB} \rho \pm}(\rho) = \frac{i}{\sqrt{2}} (\psi_{10}(\rho) \pm i \psi_{01}(\rho)).
\end{equation}
Here, $\psi_{nm}(\rho)=\psi_n(x) \, \psi_m(y)$ and $\psi_n(x)$ is the $n^{\mathrm{th}}$
harmonic-oscillator eigenstate. 
The envelope function of the admixed CB state is a superposition of 
\emph{excited} harmonic-oscillator eigenfunctions.
The prefactor 
\begin{equation}
  \lambda_{\mathrm{CB}}^\pm = \pm \frac{P}{\sqrt{2} E_g l_0}
\end{equation}
determines the degree of $sp$-hybridization. Using values from Table \ref{supp:table}
and assuming a quantum dot with dot Bohr radius $l_0 \simeq 30 \, \mathrm{nm}$ 
($\mathbf{B}=\mathbf{0}$), we estimate $\lambda_{\mathrm{CB}}^\pm \simeq 10^{-2}$.
Similarly, we estimate $\lambda_{\mathrm{LH}}^\pm \simeq \lambda_{\mathrm{SO}}^\pm \simeq 10^{-3}$,
assuming a dot height $a_z \simeq 5 \, \mathrm{nm}$.
The admixture of CB, LH and SO states to the HH state is thus on the order of 1\% and 
has therefore been neglected in our considerations.

\begin{table}
  \centering
  \begin{tabular}{|l|c||l|c|}
    \hline
    $P \, (e\mathrm{V} \mathrm{\AA})$ & 10.5$^\star$ & $\gamma'_1$ & 6.98$^\dagger$ \\
    $E_g \, (e\mathrm{V})$ & 1.52$^\star$ & $\gamma'_2$ & 2.06$^\dagger$\\
    $\Delta_{\mathrm{SO}} \, (e\mathrm{V})$ & 0.34$^\star$ & $\gamma'_3$ & 2.93$^\dagger$\\
    \hline
\end{tabular}
  \caption{Values of band parameters used in this section;
    $^\star$taken from Ref. \cite{winkler:appendix};
    $^\dagger$taken from Ref. \cite{Vurgaftman2001}.} 
    \label{supp:table}
\end{table}

We emphasize that $sp$-hybridization will lead to a coupling of the HH to
the nuclear spins via the Fermi contact interaction (\ref{ham:contact}).
Since the Fermi contact interaction is of Heisenberg-type, $sp$-hybridization will
directly lead to non-Ising corrections to the effective Hamiltonian given in 
Eq. (\ref{effham}). The size of these corrections is determined by the degree
of $sp$-hybridization which is on the order of 1\% (see above).

\section{Estimate of the Fermi contact interaction}
\label{appendix2}

In Eq. (\ref{blochHH}), we have approximated the HH $\mathbf{k}=\mathbf{0}$ Bloch
amplitudes within a Wigner-Seitz cell by a linear combination of atomic orbitals.
Similarly, we approximate the $\mathbf{k}=\mathbf{0}$ Bloch amplitude in the CB by
\begin{align}
  u_{\mathrm{CB} \mathbf{0} \sigma}(\mathbf{r}) \Bigr|_{\mathbf{r} \in \mathrm{WS}}
  = &N_{\alpha_c} \Bigl( \alpha_c \Psi_{400}^{\mathrm{Ga}}
  (\mathbf{r}+\mathbf{d}/2) \nonumber \\
  &- \sqrt{1-\alpha_c^2} \Psi_{400}^{\mathrm{As}} (\mathbf{r}-\mathbf{d}/2) \Bigr),
  \label{supp:blochCB}
\end{align}
independent of $\sigma$.
Here, $\Psi_{400}(\mathbf{r})=R_{40}(r) Y_0^0(\theta,\varphi)$, 
$\alpha_c$ describes the relative electron sharing between the Ga and As atom
in the Wigner-Seitz cell chosen to be centered halfway along the Ga-As bond, and
$N_{\alpha_c}$ normalizes the Bloch amplitude to two atoms in a primitive unit cell.
The radial wavefunction depends implicitly on the effective nuclear charges 
$Z_{\mathrm{eff}}(\kappa, 4s)$, where $\kappa = \mathrm{Ga}, \mathrm{As}$.

We will estimate the relative electron sharing in the CB by calculating the electron
densities at the sites of the nuclei from Eq. (\ref{supp:blochCB}) and comparing to
accepted values taken from Ref. \onlinecite{Paget1977}.
We will then estimate the Fermi contact interaction of a CB electron using free-atom
effective nuclear charges taken from Ref. \onlinecite{Clementi1963}
($Z_{\mathrm{eff}}(\mathrm{Ga}, 4s) \simeq 7.1$, $Z_{\mathrm{eff}}(\mathrm{Ga}, 4p) 
\simeq 6.2$, $Z_{\mathrm{eff}}(\mathrm{As}, 4s) \simeq 8.9$, and $Z_{\mathrm{eff}}(\mathrm{As}, 4p) 
\simeq 7.4$) and normalizing the Bloch amplitude over a Wigner-Seitz cell.

We approximate the electron densities at the Ga and As sites within a primitive unit cell
from Eq. (\ref{supp:blochCB}): 
\begin{align}
  \label{supp:dGa}
  d_{\mathrm{Ga}} &= |u_{\mathrm{CB} \mathbf{0} \sigma}(-\mathbf{d}/2)|^2
  \simeq N_{\alpha_c}^2 \alpha_c^2 |\Psi_{400}^{\mathrm{Ga}}(\mathbf{0})|^2,\\
  d_{\mathrm{As}} &= |u_{\mathrm{CB} \mathbf{0} \sigma}(+\mathbf{d}/2)|^2
  \simeq N_{\alpha_c}^2 (1-\alpha_c^2) |\Psi_{400}^{\mathrm{As}}(\mathbf{0})|^2.
    \label{supp:dAs}
\end{align}
We estimate the corrections to the right-hand sides to be on the order of 1\% due to overlap terms.
We take the ratio $d_{\mathrm{Ga}}/d_{\mathrm{As}}$ and equate this with the ratio of the 
values from Ref. \onlinecite{Paget1977},
$d_{\mathrm{Ga}}' = 5.8 \times 10^{-31} \> \mathrm{m}^{-3}$ and
$d_{\mathrm{As}}' = 9.8 \times 10^{-31} \> \mathrm{m}^{-3}$.
This allows us to write $\alpha_c$ as a function of the two effective nuclear charges:
\begin{equation}
  \label{supp:beta}
  \alpha_c = \left[ 1 +  \frac{d_{\mathrm{As}}'}{d_{\mathrm{Ga}}'}
  \left( \frac{Z_{\mathrm{eff}}(\mathrm{Ga},4s)}{Z_{\mathrm{eff}}(\mathrm{As},4s)} \right)^3 \right]^{-1/2}.
\end{equation}
Recalling that $N_{\alpha_c}$ normalizes the Bloch amplitude to two atoms over a Wigner-Seitz cell, 
we write
\begin{align}
  N_{\alpha_c} = \biggl[ \frac{1}{2} \int_{\mathrm{WS}} &d^3 r \> \Bigl| \alpha_c
  \Psi_{400}^{\mathrm{Ga}}(\mathbf{r}+\mathbf{d}/2) \nonumber\\
  &-\sqrt{1-\alpha_c^2} \Psi_{400}^{\mathrm{As}}(\mathbf{r}-\mathbf{d}/2) \Bigr|^2 
  \biggr]^{-1/2}. \label{supp:Nbeta}
\end{align}
For all numerical integrations, we approximate the Wigner-Seitz cell by a sphere
centered halfway along the Ga-As bond
with radius equal to half the Ga-Ga nearest-neighbor distance.
Inserting (\ref{supp:beta}) and (\ref{supp:Nbeta}) into (\ref{supp:dGa}) and
(\ref{supp:dAs}), we solve the two coupled equations
\begin{align}
  d_{\mathrm{Ga}}(Z_{\mathrm{eff}}(\mathrm{Ga}), Z_{\mathrm{eff}}(\mathrm{As})) - 
  d_{\mathrm{Ga}}' &= 0,\\
  d_{\mathrm{As}}(Z_{\mathrm{eff}}(\mathrm{Ga}), Z_{\mathrm{eff}}(\mathrm{As})) - 
  d_{\mathrm{As}}' &= 0,
\end{align}
for the two effective nuclear charges. This yields ${Z_{\mathrm{eff}}(\mathrm{Ga}) \simeq 9.8}$ 
and ${Z_{\mathrm{eff}}(\mathrm{As}) \simeq 11.0}$.
Inserting these values back into Eq. (\ref{supp:beta}), we estimate the electron sharing 
within the primitive unit cell to be 
\begin{equation}
  \alpha_c^2 \simeq 0.46.
\end{equation}
For comparison, inserting free-atom effective nuclear charges into Eq. (\ref{supp:beta})
yields a similar value: $\alpha'^2_c \simeq 0.54$.

Now we estimate the Fermi contact interaction of a CB electron starting from the free-atom
effective nuclear charges $Z_{\mathrm{eff}}(\kappa,4s)$ 
obtained from Ref. \onlinecite{Clementi1963}. We use $\alpha_c \simeq 1/\sqrt{2}$ and normalize the
$\mathbf{k}=\mathbf{0}$ Bloch amplitude to two atoms over a Wigner-Seitz cell,
following Eq. (\ref{supp:Nbeta}). From the normalized Bloch amplitudes we estimate
the Fermi contact hyperfine interaction by evaluating
\begin{equation}
  A_e^j = \frac{2 \mu_0}{3} \> \gamma_S \gamma_{j} |u_{\mathrm{CB} \mathbf{0} \sigma}(\mathbf{R}_j)|^2.
\end{equation}
Here, $\mathbf{R}_j = \mp \mathbf{d}/2$ for Ga and As, respectively ($j$ indexes the
nuclear isotope).
Evaluating for the isotopes in GaAs, this gives the values shown in column (iii) of Table 
\ref{table1}.

Replacing the Wigner-Seitz cell by a sphere with radius $R_s$ equal to half the Ga-Ga 
nearest-neighbor distance in our numerical integrations overestimates the expectation 
value of $[r_k^3 (1+d/r_k)]^{-1}$ in Eq. (\ref{holecoupling}).
To estimate the error, we perform an integration over a sphere with radius 
$R'_s=(R_s+R_{\mathrm{max}})/2$, where $R_{\mathrm{max}}$ denotes the radius of the smallest sphere that
fully contains the Wigner-Seitz cell. From this, we estimate the relative error to be 
less than 30\%.

\section{Estimate of the long-ranged interactions}
\label{appendix3}

In this section, we estimate the corrections to the HH coupling strength in Eq. (\ref{holecoupling})
due to long-ranged dipole-dipole interactions and long-ranged
$\mathbf{L} \cdot \mathbf{I}$ interactions. To this end, we consider a single
nucleus interacting with a HH that is delocalized over the lattice sites in the quantum dot.
We start from the Hamiltonians given in Eqs. (\ref{ham:anisotropic}) and (\ref{ham:angular}).
We define effective radii $a_{\mathrm{eff}}(\kappa,4p) = a_0/Z_{\mathrm{eff}}(\kappa,4p)$,
where $a_0 \simeq 5.3 \times 10^{-11} \, \mathrm{m}$ is the Bohr radius. The effective radii 
define an approximate length scale for the spread of the site-localized functions
$\Psi_{41\sigma}^\kappa(\mathbf{r})$ and are much smaller than the GaAs lattice constant
$a_L \simeq 5.7 \times 10^{-10} \, \mathrm{m}$. The nucleus thus
effectively `sees' sharp-peaked electron densities centered around the more distant lattice sites.
We choose the nucleus to be at site $\mathbf{R}_k$ and estimate the interaction
with the electron density at more distant atomic sites 
by approximating the electron densities by $\delta$-functions.
Adding up contributions from $h_2^k$ and $h_3^k$, we arrive at an effective Hamiltonian
describing the long-ranged interactions: $H^k_{\mathrm{lr}} = A^k_{\mathrm{lr}} s_z I_k^z$,
where $A^k_{\mathrm{lr}} = \sum_{l; l \neq k} A^{kl}_{\mathrm{lr}}$ is the associated coupling strength
and $A^{kl}_{\mathrm{lr}} = v_0 |\phi(\mathbf{R}_l)|^2 \int d^3 r_k \{ \delta(\mathbf{r}_k
-\mathbf{R}_{kl}) (h_2^k+h_3^k) \}$ describes the coupling 
of the electron density at site $\mathbf{R}_l$ to the nucleus at site $\mathbf{R}_k$
($\mathbf{R}_{kl} = \mathbf{R}_l - \mathbf{R}_k$). In order to estimate the
size of the long-ranged interactions relative to the on-site interactions, we take into
account nearest-neighbor couplings for the long-ranged part, i.e., we replace $\sum_{l;l \neq k} \rightarrow
\sum_{l = \mathrm{n.n.}}$. The interaction with electron
density located around more distant nuclei is suppressed by $\sim 1/R_{kl}^3$.
Assuming that the quantum-dot envelope function varies slowly over the
nearest-neighbor distance ($\phi(\mathbf{R}_l) \simeq \phi(\mathbf{R}_k)$ for $l$ nearest neighbor
of $k$), we estimate the ratio of long-ranged and on-site interactions 
(Table I, column (i)) to be
\begin{equation}
  \label{supp:lrestimate}
  \frac{A_{\mathrm{lr}}}{A_h} \simeq 7 \times 10^{-3},
\end{equation}
on the order of 1\%, where $A_{\mathrm{lr}}^k = A_{\mathrm{lr}} v_0 |\phi(\mathbf{R}_k)|^2$.

We remark that, in principle, the electron g-factor can deviate from the free-electron 
g-factor due to spin-orbit interaction. According to Ref. \onlinecite{Yafet1961},
this renormalization is negligible for the on-site interaction, but could become
relevant for the long-ranged interaction. However, for the estimate in Eq. 
(\ref{supp:lrestimate}), we have taken the free-electron g-factor.

\section{Variance of the nuclear field}
\label{appendix4}

Here we calculate the nuclear-field variance for a HH interacting
with nuclei in a quantum dot. In particular, we evaluate
\begin{equation}
  \sigma^{2}=\left\langle h_{z}^{2}\right\rangle,
\end{equation}
where $\langle \cdots \rangle =\mathrm{Tr}_{I}\left(\bar{\rho}_{I}\cdots\right)$
indicates the expectation value with respect to the infinite-temperature
thermal equilibrium density matrix $\bar{\rho}_{I}$ and we recall
$h_{z}=\sum_{k}A_{k}^{h}I_{k}^{z}$. For an uncorrelated and unpolarized
nuclear state, we have $\left\langle I_{k}^{z}I_{k^{\prime}}^{z}\right\rangle 
=\left\langle I_{k}^{z}\right\rangle \left\langle I_{k^{\prime}}^{z}\right\rangle =0,\; k\ne k^{\prime}$,
which gives
\begin{equation}
  \sigma^{2}=\sum_{k}\left(A_{k}^{h}\right)^{2} \langle (I_{k}^{z})^{2} \rangle .
\end{equation}
Using ${\langle (I_{k}^{z})^{2} \rangle =I^{j_{k}}\left(I^{j_{k}}+1\right)/3}$
for an infinite-temperature state, $A_{k}^{h}=A_{h}^{j_{k}}v_{0}|\phi(\mathbf{R}_{k})|^{2}$,
and assuming that the nuclear isotopic species with abundances $\nu_{j}$
are distributed uniformly throughout the dot gives
\begin{equation}
  \sigma^{2}=\frac{1}{3}I_{0}\sum_{j}\nu_{j}I^{j}(I^{j}+1) (A_{h}^{j})^{2},
  \label{supp:SigmaSquaredDefinition}
\end{equation}
where
\begin{equation}
  I_{0}=v_{0}^{2}\sum_{k}|\phi(\mathbf{R}_{k})|^{4}.
  \label{supp:I0Definition}
\end{equation}
Assuming that the envelope function $\phi(\mathbf{r})$ varies slowly
on the scale of the lattice, we replace the sum in Eq. (\ref{supp:I0Definition})
by an integral: \begin{equation}
v_{0}\sum_{k}|\phi(\mathbf{R}_{k})|^{4}\to\int d^{3}r|\phi(\mathbf{r})|^{4}.\label{supp:SumToIntegral}\end{equation}

Inserting the envelope functions (\ref{supp:enveloperho}) and (\ref{supp:envelopez}) 
for a quantum dot with height $a_{z}$ and radius $l$ and evaluating the integral in Eq. 
(\ref{supp:SumToIntegral}), we find
\begin{equation}
  I_{0}=\frac{3}{4}\frac{1}{N}.\label{supp:I0Evaluated}
\end{equation}
Here, $N$ is the number of nuclear spins within the quantum dot,
given explicitly by \begin{equation}
N=\frac{\pi l^{2}a_{z}}{v_{0}}.\end{equation}
Inserting Eq. (\ref{supp:I0Evaluated}) into Eq. (\ref{supp:SigmaSquaredDefinition})
directly gives Eq. (\ref{sigma}).

\end{appendix}


\end{document}